# De-Fragmenting the Cloud

Mayank Mishra
Indian Institute of Technology Bombay
Mumbai, India
mayank@cse.iitb.ac.in

Umesh Bellur
Indian Institute of Technology Bombay
Mumbai, India
umesh@cse.iitb.ac.in

## ABSTRACT

Existing VM placement schemes have measured their effectiveness solely by looking either Physical Machine's resources(CPU, memory) or network resource. However, real applications use all resource types to varying degrees. The result of applying existing placement schemes to VMs running real applications is a fragmented data center where resources along one dimension become unusable even though they are available because of the unavailability of resources along other dimensions. An example of this fragmentation is unusable CPU because of a bottlenecked network link from the physical machine which has available CPU. To date, evaluations of the efficacy of VM placement schemes has not recognized this fragmentation and it's ill effects, let alone try to measure it and avoid it. In this paper, we first define the notion of what we term "relative resource fragmentation" and illustrate how it can be measured in a data center. The metric we put forth for capturing the degree of fragmentation is comprehensive and includes all key data center resource types. We then propose a scheme of minimizing this fragmentation so as to maximize the availability of existing set of data center resources. Results of empirical evaluations of our placement scheme compared to existing network based placement schemes show a reduction of fragmentation by as much as 15% and increase in number of successfully placed applications by upto 20%.

## 1. INTRODUCTION

Virtual Machine (VM) placement is one of the crucial operations in data center as it directly affects data centers efficiency and performance. VM placement has been proven to be one of the most challenging operations in data center as well, as it has been shown that optimal VM placement is NP-Hard [26, 16]. Hence, existing VM placement schemes simplify the problem by considering only a subset of resources, for example, considering either network bandwidth requirement between VMs [13, 15] or CPU and memory requirement of VMs [25, 19] though applications use all resource types to varying degrees.

In general, a VM placement plan which has been optimized for utilization of a subset of resources may not necessarily result in efficient utilization of other resources. Hence, employing existing placement schemes, which optimizes either physical machine's resources or network resource, may result in a fragmented data center where available resources along one dimension become unusable due to the unavailability of resources along other dimensions. An example of this fragmentation is unusable available CPU due to a bottlenecked network link [4, 18, 13]. We present this fact by using a simple toy example.

Figure 1 shows a simple scenario where three communicating VM-pairs (A1-A2, B1-B2, C1-C2) are to be placed on two hosts, Host1 and Host2, connected by a switch. The CPU requirements of VMs are shown as percentage of host machine's CPU capacity. Similarly, the network bandwidth requirement of VM pairs is shown as percentage of link bandwidth capacities. For simplicity, only CPU and network bandwidth are the considered. One very commonly used placement scheme for VM placement, which has been adopted from single dimensional bin packing [1], is "First Fit Decreasing" (FFD), where objects are first sorted in decreasing order of their sizes and then placed on bins in first fit manner. For the sorting of VMs, the size is represented either by the most dominating resource [19, 25, 14] or by combination of resources [26, 16, 22]. Figure 1 shows two sorted orders at the top.

In the example, VMs are first placed on hosts using FFD heuristic considering only CPU as shown in Plan (A). However, after placement, the network bandwidth requirement between Host1 and Host2 can not be fulfilled as requirement is 115 units wheareas the capacity is 100. This leads to under-performance of the hosted VMs. [4, 18] discuss the adverse effect of network unavailability on application performance. In data centers where an explicit reservation of bandwidth is required, the placement is not possible. Plan (B) in the figure shows VM pairs placed according to decreasing order of bandwidth requirement (by co-locating the communicating VMs). It can be seen that the CPU capacity of Host2 is exceeded. Thus, again, the performance of VMs will be degraded and in reservation based schemes the placement is not possible. In Plan (A), there was sufficient CPU capacity however it remained unutilized due to unavailability of network bandwidth and in the Plan (B), because of unavailable CPU capacity, the network bandwidth remained unutilized. There is indeed a valid and successful placement possible and is shown in case C. This problem of unusablilty of a resource due to unavailability of other resource is referred by the name "Relative Resource Fragmentation (RRF)" problem.

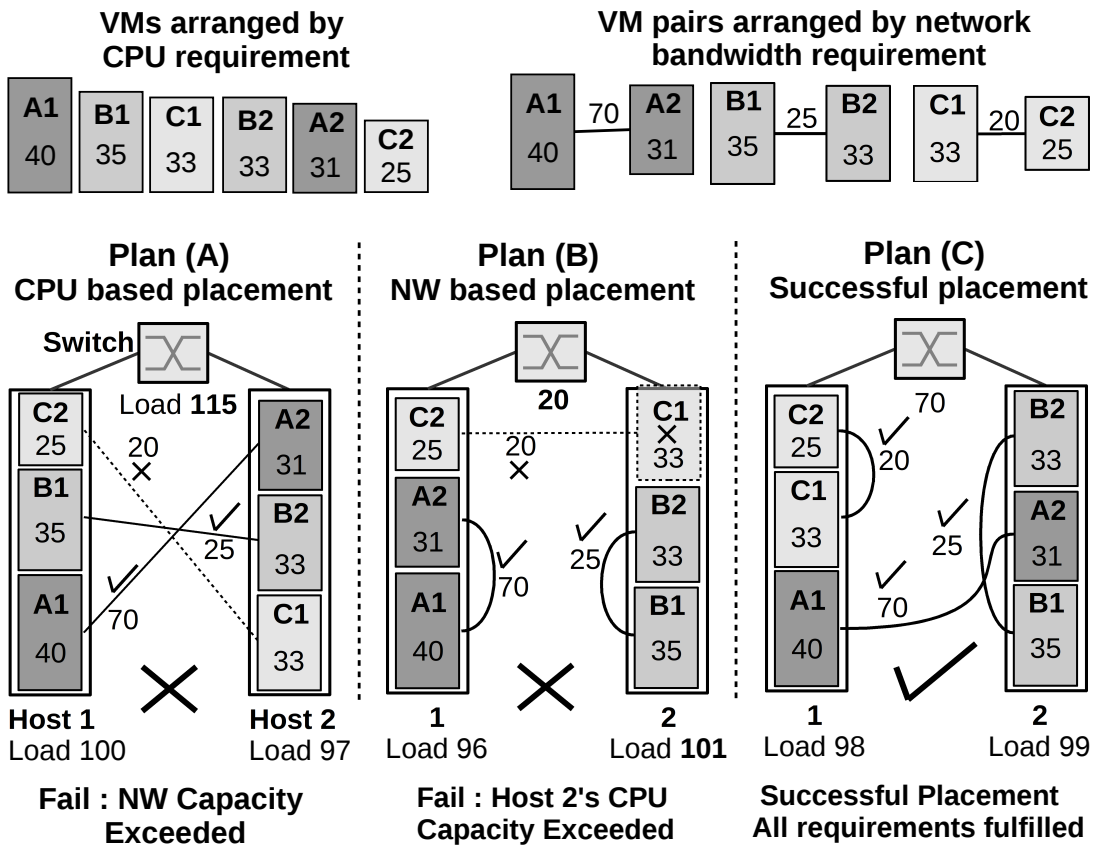

**Figure 1: VM Placement involving CPU, network**

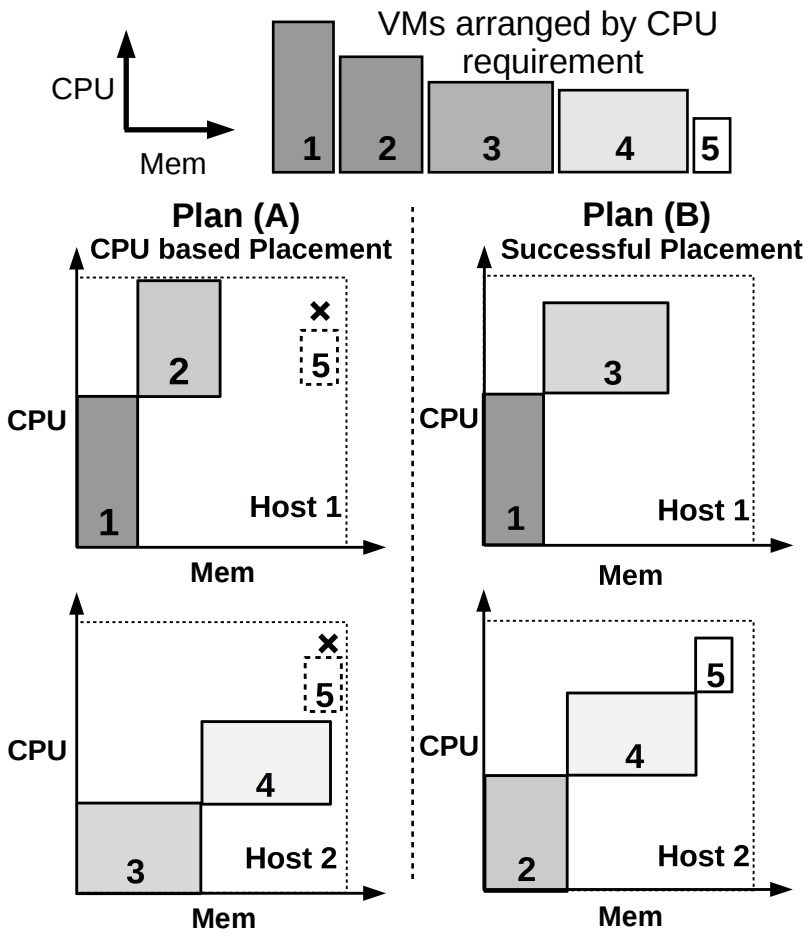

**Figure 2: VM Placement involving CPU, Mem**

The problem of RRF is independent of the resources involved. For example, Figure 2 shows a scenario where 5 VMs, which need certain amount of CPU and memory, are required to be placed (network is not considered). Plan (A) shows what happens when only CPU requirement is considered for VM placement using FFD based placement scheme. The fifth VM can not be placed even though collectively, in both PMs combined, there is adequate CPU and memory. PM1 has sufficient amount of memory but not CPU and vice-versa for PM2.

RRF refers to a condition when one or more of the available resources of a section of data center remains underutilized (or unutilized) because of the unavailability of one or more other resources. Such conditions result in underutilization of data center and potentially causes revenue loss. To date, evaluations of the efficacy of VM placement schemes has not recognized this fragmentation and it's adverse effects, let alone try to measure it and avoid it. For example [4, 18, 13] discuss unpredictability in the performance of applications and low utilization of CPU when network Bandwidth requirement between VMs is not considered for placement decisions. They try to solve the problem by proposing only network based placement approaches but fail to consider the effect of these new placement approaches on utilization levels of CPU. To simplify the problem they assume the existence of CPU/memory slots (fixed size shares) and VMs needs 1 slot each. In this paper we contend and show that such approaches increase the problem RRF in the data centers. Our goal is to design a VM placement scheme which considers all datacenter resource types and minimizes the RRF. The major contributions of this paper are:

1. Proposed a comprehensive metric for capturing the datacenter fragmentation called RRF and illustration of how it can be measured in a data center in Section 2.

2. Proposed a novel notion of "`Reach`" which is used to find achievable network bandwidth in a data center (which simplifies RRF measurement and application placement) in Section 3.

3. A unified application placement scheme to minimize RRF so as to maximize the availability of existing set of data center resources is proposed in Section 5. This unified application placement scheme considers all the resources present in data center.

4. Results of empirical evaluations mention in Section 6 using the real world datasets [2, 5] show that our VM placement scheme results in reduction of RRF by as much as 15% and is able to successfully place upto 20% more applications.

This paper has two parts- Sections 2 and 3 discuss about RRF and its calculation and Sections 4 and 5 propose the VM placement scheme to reduce RRF in datacenters.

## 2. RESOURCE FRAGMENTATION PROBLEM

Fragmentation refers to a condition where the availability of a resource is scattered making it unusable for a allocation request. Fragmentation level of a resource is an indicator of the quality of resource utilization [11]. Quantification of fragmentation, which is mostly associated with memory and storage resources, has been an active area of discussion.

RRF, on the other hand, refers to a situation where one or more of the available resources remains underutilized (or unutilized) because of the unavailability of other resources. In other words, RRF refers to fragmentation in multiple resource dimensions.

We consider the existence of $k$ different resource types $r_1, r_2, \ldots, r_k$. For simplicity, we consider $k = 3$, where, the considered resources are CPU, memory and network bandwidth. Other important data center resources can be easily incorporated into the framework. Please note that the fragmentation and RRF of a resource are always quantified with respect to a resource requirement given in the request.

Formally, allocation request for resource $r_i$ for normalized size $s_i$ relative to which fragmentation needs to be calculated is denoted by $Req(r_i, s_i)$ . In the case of CPU and memory, the request is normalized to the capacity of a single host and in the case of network bandwidth, the request is normalized to the capacity of a single link. For example, $Req(cpu, 0.2)$ denotes that 20% of host's CPU capacity is requested.

Next, we present formal definitions of fragmentation and RRF in data centers and then propose mathematical formulas to quantify them.

## 2.1 Fragmentation Index

The fragmentation Index for a resource $r_i$ for a given request $Req(r_i, s_i))$ (where, $1 \leq i \leq k$) is denoted by $F(r_i, Req(r_i, s_i))$. It is the ratio of the aggregate of fragmented remaining resource capacities of the hosts where resource request $Req(r_i, s_i)$ can not be allocated to the total free capacity of resource $r_i$ across all hosts. Formally, the fragmentation index of a resource $r_i$ can be the seen as the following ratio

$$\left( \frac{\text{Aggregate free capacity of } r_i \text{ which can not be occupied by a request}}{\text{Total available free capacity of } r_i} \right) \quad (1)$$

Mathematically it can be expressed as

$$F(r_i, Req(r_i, s_i)) = \frac{T(r_i) - N \times s_i}{T(r_i)} \quad (2)$$

where, $T(r_i)$ which denotes the aggregate total free capacity of resource $r_i$ across all hosts and $N$ is the number of requests which can be simultaneously fulfilled.

Fragmentation index ranges from 0 to 1. Fragmentation index of 0 is the best scenario as all the available capacity is usable. Fragmentation index of 1 means that no usable capacity is available.

**Example:** Figure 3 shows an example where 3 hosts have certain level of CPU and memory utilizations. The calculation of memory fragmentation index for a normalized request of 0.25 (denoted by $F(mem, Req(mem, 0.25)))$ is shown below. The observed fragmentation index is 0.166.

$$\begin{aligned} T(mem) &= 0.2 + 0.5 + 0.5 = 1.2 \\ N &= 4 \\ F(mem, 0.25) &= \frac{1.2 - 4 \times 0.25}{1.2} = 0.166 \end{aligned}$$

Similarly, memory fragmentation index for a normalized request of 0.3 is 0.5.

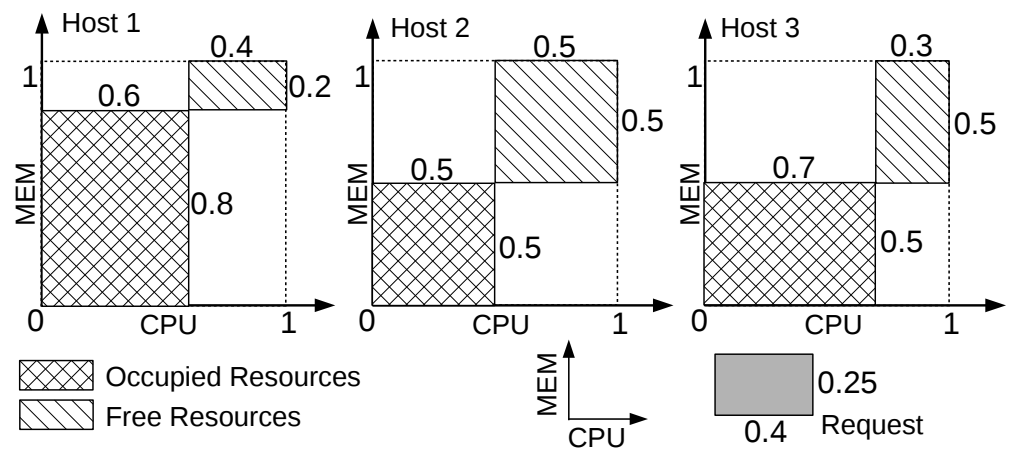

**Figure 3: Calculating CPU's and memory's Fragmentation and RRF**

## 2.2 RRF Index

RRF Index of a resource $r_i$ for given multidimensional request $(Req(r_1, s_1),\ Req(r_2, s_2), ..., Req(r_k, s_k))$ (where, $1 \leq i \leq k$) is the ratio of the aggregate of fragmented remaining resource capacities of $r_i$, where multidimensional resource request, can not be fulfilled, to the total aggregate remaining capacity of resource $r_i$.

RRF index of 0 denotes that all the available capacity is usable and RRF index of 1 means that no usable capacity is available. For calculating RRF index the request should be of more than one dimension.

Formally RRF index of a resource $r_i$ can be the seen as the following ratio

$$\left( \frac{\text{Aggregate free capacity for } r_i \text{ which can not be occupied by a multidimensional request}}{\text{Total available free resource capacity for } r_i} \right) \quad (3)$$

Thus, $RRF(r_i, Req(r_1, s_1), Req(r_2, s_2), ..., Req(r_k, s_k))$ is given by

$$\frac{T(r_i) - N_m \times s_i}{T(r_i)} \quad (4)$$

where $N_m$ is the number of multi-dimensional requests which can be simultaneously fulfilled.

**Example:** In the example shown in Figure 3, the RRF index of memory for request $[Req(mem, 0.25), Req(cpu, 0.4)]$ is calculated using Equation 4. It can be seen that number of such requests which be simultaneously fulfilled $N$ is 1. Thus, $RRF(mem, Req(mem, 0.25), Req(cpu, 0.4))$ is calculated as

$$\frac{1.2 - 1 * 0.25}{1.2} = 0.791$$

Fragmentation and RRF manifest themselves in terms of the number of requests which can be simultaneously fulfilled. For simplicity, in the rest of this paper, we will consider only the number of placeable requests as the measure of Fragmentation and RRF. Lower the Fragmentation or RRF index, higher the number of requests which can be placed simultaneously.

## 2.3 RRF Index for Network

Calculation of network RRF index, is more complex than CPU and memory. The reason is that network bandwidth is always between a pair of communicating end points. Thus, the allocation request for calculating the RRF must consider *cpu*, *mem* requirements of both the end points on all possible pairs of hosts. We only show the calculation for network RRF and not fragmentation because calculation of fragmentation index can easily be derived from the calculation of RRF index by reducing the number of dimensions of resource request. We now show the network RRF index calculation. Consider a simple data center with four hosts H1 to H4 and 3 switches S1 to S3 shown in Figure 4. First, we calculate the total available network capacity of the datacenter, then we discuss the representation of multidimensional resource allocation request. Then, we count the number of placeable multidimensional requests and using Equation 4 we calculate the network RRF index.

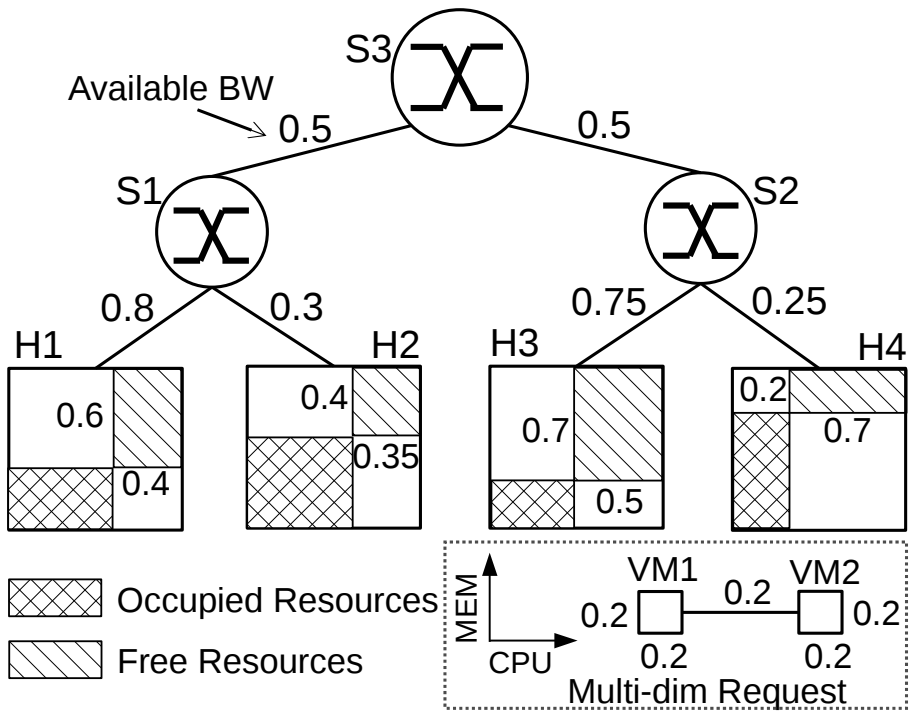

**Figure 4: Calculating network RRF**

**Calculating total achievable network capacity** $T(nw)$**:** The first step to calculate network RRF is to find $T(nw)$. As the network bandwidth is always utilized between two end points, there are many possible pairs of hosts to consider in figure 4. On path H1-S1-H2, available bandwidth is 0.3. On path H3-S2-H4, available bandwidth is 0.25, On path H1-S2-S3-H3, available bandwidth is 0.5. Thus, $T(nw) = 0.3{+}0.25{+}0.5 = 1.05$. It should be noted that once a certain amount the bandwidth on a link is considered between a pair of hosts, it cannot be considered for any other host pair.

**Representing Request:**To calculate network RRF index, the multidimensional request is represented as $[CPU_{src}, Mem_{src}, NW_bw_{src,dst}, CPU_{dst}, Mem_{dst}]$. This multidimensional request is shown in Figure 4. For simplicity, the source and destination end points are assumed to have same requirement of CPU and memory and hence the request can be represented as $[Req(cpu, 0.2), Req(mem, 0.2), Req(nw, 0.2)]$.

**Number of satisfiable requests:** Between H1 and H3, two requests can be placed as there are adequate amount of resources on both H1 and H3 and on path between them. Once these two requests are placed between H1 and H3, no more requests can be placed on this data center. Though, there is enough CPU and memory available on H2 and H4, we can not place any more requests due to unavailability of network bandwidth. But, is this the maximum possible requests which can be placed? The answer is No.

Consider placing one request each between H1, H2 and H3, H4. Once these requests are placed there will still be enough capacity left in terms of CPU, memory and network between H1 and H3 to place one more request. Thus, three requests can be placed. Using Equation 4, we calculate network RRF index of the example mentioned in Figure 4 to be

$$\frac{1.05 - 3 * 0.2}{1.05} = 0.428$$

An important question which arises is "what should be the strategy to count the number of placeable requests so that maximum number of requests can be accounted?". For smaller topologies, like the one shown in example, it is easy to count and verify the maximum number of placeable requests but in real world data center topologies containing thousands of hosts and hundreds of switches the problem of counting becomes non-trivial. To solve this counting problem we need to first understand why the number of placeable requests differ in the way they are counted. The reason of such difference is presented in next sections (Section 3, 3.1). The algorithm to count the number of placeable requests is also presented in Section 3.3.

# 3. MAXIMUM CAPACITY OF DATA CENTER NETWORK

The network capacity of data center depends on where the communicating entities are placed. For example, consider two VMs A and B which have a certain communication bandwidth requirement between them. There are three ways to place these VMs - **1) Place on the same host:** The network infrastructure of the data center remains unused. Theoretically, there is infinite network capacity as no network infrastructure is utilized. **2) Place on hosts in the same rack (connected to same switch):** Only 2 links (from host A to switch and from switch to Host B, i.e., 2 hops) are used. In this case the achievable network capacity is not infinite but half of the aggregate capacities of all the links between hosts and their switches (because of 2 links or hops used per communicating pair). **3) Place on different racks:** More than 2 links are utilized. The achievable network capacity is much lesser than "same rack" case. The worst case occurs when all the communication passes through min-cut (or bisection) of the network topology graph. Thus the placement of communicating VM pair is crucial in deciding the achievable network capacity of a data center. The key point is to place the communicating VMs as close to each other as possible. However, practical constraints of resource availability, host and rack resource capacities result in placements which are not able to exploit the full capacity of data center network. It can be seen that full network capacity of data center refers to the bandwidth available when all the pairs of communicating VMs are placed on hosts connected to the same switch, i.e., placed on same rack.

The question which arises is whether communicating endpoints are always required to be connected to same switch to exploit full capacity of data center or there are some relaxations possible? The answer lies in the data center's topological characteristic namely bisection (or bisection) bandwidth. In the next section we discuss the characteristics of data center network infrastructure which can be exploited to relax the constraint of two hop distance between communicating endpoints for maximum achievable data center network capacity.

## 3.1 Data center Network Topology Characteristics - The reach

Network topologies of data centers rarely have full bisection bandwidth[1] across them [18, 4, 13, 8, 6]. data center topologies are generally oversubscribed (or have low bisection bandwidth) towards the core level switches as shown in Figures 5(a), (b). However, at non core levels, certain regions (or subgraphs) of full bisection bandwidth do exist. For example, a portion of topology consisting of a "Top of the Rack Switch (TOR)" and hosts connected to it does have full bisection bandwidth. In this paper, such regions (or subgraphs) with full bisection bandwidth are referred as the "`reach`" of the hosts. Different topologies have different sizes of the "`reach`". For example, in a simple tree topology, there are as many `reaches` as TORs, i.e., every TOR along with its hosts is a `reach` as shown in Figure 4, whereas, in a Fat Tree, the complete topology is one `reach` (refer Figure 5(c)).

Inside a `reach`, due to full bisection bandwidth, the relative location of communicating VMs does not have any effect on bandwidth availability. Thus, from network bandwidth availability point of view, two communicating end points located anywhere in the `reach` are as "close" to each other as if they are located on the same rack. The concept of `reach` is used to find- **1)** total achievable network capacity $T(nw)$ and **2)** the number of placeable requests $N_m$ for calculating RRF index of network. Inside a `reach`, only host's resources like CPU, memory and NIC capacity need to be considered for counting the number of placeable requests as the network bandwidth between the end points is always available due to full bisection bandwidth. The total achievable network capacity $T(nw)$ is calculated as the sum of achievable network capacity inside the `reaches` and then the achievable capacity between the `reaches`. The exact steps to calculate network capacities inside and between `reaches` is proposed in Section 3.3. Next, we briefly discuss the method of finding `reaches` in topology.

[1] bandwidth across smallest cut that divides network into two equal parts

## 3.2 Finding reaches in topology

We now propose a simple scheme to find `reach` in tree based topologies. Certain assumptions are a) host is connected to only 1 TOR, b) there are even number of hosts on a rack and c) there are even number of children for each parent node. A `reach` consists of a set of hosts and the set of switches which are further connected to other part of topology via an oversubscribed network link. It can be seen from Figure 5 that the "`reaches`" always lie in non-oversubscribed zone. For ease of explanation, call the switches beyond which there is oversubscription in network links as "boundary-switches". For example, in Figure 5 all the TOR switches in a tree topology and all aggregator switches in CLOS topology are boundary switches. Let $S$ be the set of all boundary-switches. The key intuition behind `reach`-finding algorithm is to divide $S$ into subsets such that all member boundary-switches in subset share common hosts. We now propose the `reach`-finding algorithm. Let $R$ denote the set of `reaches`, initialize $R$ to empty set ($R \leftarrow \varnothing$).

**Find Reaches**

1. Let $S_v$ denotes visited switches. Initialize $S_v$ ($S_v \leftarrow \varnothing$.
2. For every switch $s \in S$ do
   (a) If $s \in S_v$ then go to 2
   (b) Let $H$ denote the hosts which have $s$ in their parent chain.
   (c) Let $P$ denote switches which are in parent chain of hosts in $H$ at same level as $s$.
   (d) Let $r$ be the new `reach`. $r.hosts \leftarrow H$, $r.switches \leftarrow P$.
   (e) $S_v \leftarrow S_v \cup P$
   (f) $R.add(r)$
3. $R$ gives the set of `reaches`

We now discuss the algorithm to find network RRF in a data center.

## 3.3 Network RRF Calculation Algorithm

First, the steps to find total achievable network capacity $T(nw)$ are proposed. Note that the achievable capacity of the network is dependent on the way it is utilized as discussed before. There are two components of $T(nw)$ **a)** achievable network capacity inside the `reaches` denoted by $T_R(nw)$ and **b)** achievable network capacity between the `reaches` denoted by $T_{BR}(nw)$. Following procedures explain the way these capacities are calculated. Please note that these capacities can be used both in network fragmentation and RRF calculation.

**A) $T_R(nw)$: Achievable Network Capacity Inside `reaches`**

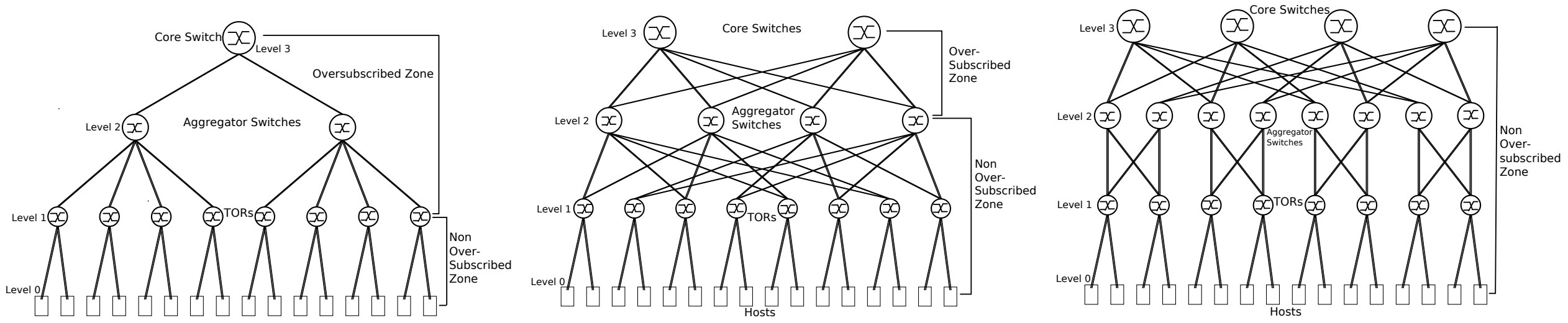

(a) Tree Topology : Bijection Bandwidth = 1

(b) CLOS Topology : Bijection Bandwidth = 4

(c) Fat Tree Topology : Bijection Bandwidth = 8

**Figure 5: Datacenter topologies with different bisection bandwidths (in terms of links)**

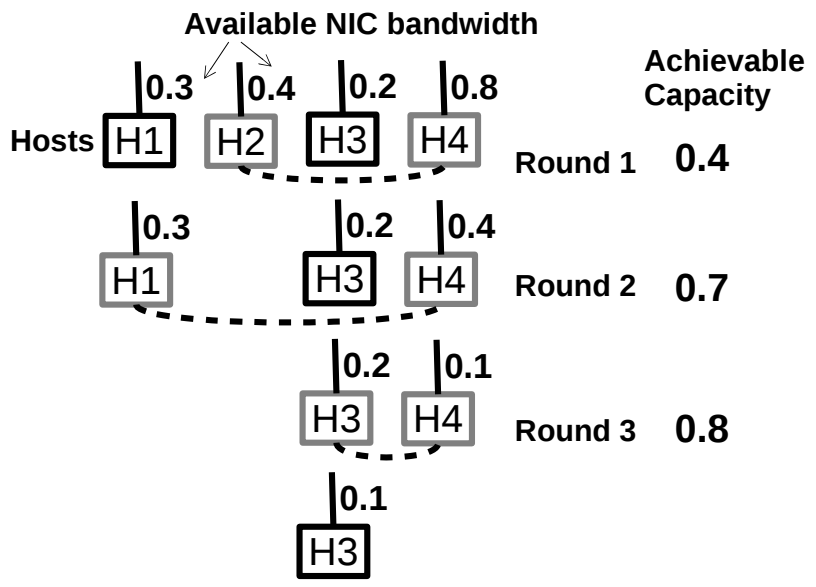

**Figure 6: Network capacity inside `reach`**

1. Initialize achievable bandwidth. $B \leftarrow 0$.
2. Let the set of `reaches` be $R$.
3. For every `reach` $r \in R$, repeat steps 4 to 8
4. Let $H$ denotes the set of all hosts in the current `reach`.
5. From $H$, find two hosts which have maximum and second maximum available NIC capacity, call them $h_{max}$ and $h_{s_max}$ and their available NIC capacities as $nic_{max}$ and $nic_{s_max}$ respectively.
6. If pair is found then
   (a) Increase the achievable bandwidth,i.e., $B \leftarrow B + nic_{s_max}$.
   (b) Reduce available NIC capacities of $h_{max}$ and $h_{s_max}$, i.e., $nic_{max} \leftarrow nic_{max} - nic_{s_max}$ and $nic_{s_max} \leftarrow 0$).
   (c) Remove $h_{s_max}$ from $H$.
7. If more than one host remaining in $H$ then goto step 5.
8. Available NIC capacity $nic_{last}$ of the last host $h_{last}$ is assigned as residual capacity to the `reach` $r$. $r.res_bw \leftarrow nic_{last}$
9. $B$ gives the achievable bandwidth inside `reaches`.

Figure 6 gives the key intuition behind calculation of $T_R(nw)$.

### B) $T_{BR}(nw)$: Achievable network capacity between `reaches`

1. Initialize achievable bandwidth. $B \leftarrow 0$.
2. List all possible pairs of `reaches` in a set $L$. Note that $|L| = |R| \times |R|$.
3. Repeat steps 4 to 9 till set $L$ is not empty.
4. Select a `reach`-pair from $L$, say $(r_i, r_j)$, where `reaches` have minimum path length between them. Note that a reach pair can have multiple paths of same length between them. In case of a tie, select the `reach`-pair where `reaches` have maximum available bandwidth between them. Let $BW(r_i, r_j)$ denotes the bandwidth between pair of `reaches`.
5. Let $B_{min}$ denotes the achievable network capacity between the `reach`-pair $(r_i, r_j)$. Here, $B_{min} = min(d_{r_i}, d_{r_j}, BW(r_i, r_j))$.
6. Reduce available bandwidth between `reaches` $r_i$ and $r_j$ by $B_{min}$.
7. Reduce the residual bandwidth associated with the `reaches`. $r_i.res_bw \leftarrow r_i.res_bw - B_{min}$ and $r_j.res_bw \leftarrow r_j.res_bw - B_{min}$.
8. Increase the achievable bandwidth, i.e., $B \leftarrow B + B_{min}$
9. Remove the `reach`-pair from $L$. $L \leftarrow L - (r_i, r_j)$.
10. $B$ gives the achievable network capacity between `reaches`.

The total available bandwidth is given by $T(nw) = T_R(nw) + T_{BR}(nw)$. Similar to the total achievable bandwidth, the number of placeable requests also needs to be calculated separately for inside-`reaches` and between-`reaches` portions of the topology. Let the multidimentional request be given as
$[Req(nw, s_n), Req(cpu, s_c), Req(mem, s_m)]$

### $N_m$: Number of placeable requests inside `reaches`

1. Initialize number of placeable requests, i.e., $N_m \leftarrow 0$.
2. For every `reach` $r, r \in R$, repeat steps 4 to 8.
3. Let $H$ denote the set of Hosts in $r$ which have available NIC bandwidth greater or equal to $s_m$.
4. For every host $h \in H$,
   (a) let $v_c, v_m, v_n$ denote number of CPU request of size $s_c$, memory request of size $s_m$ and NIC bandwidth request of size $s_n$ respectively which host $h$ can satisfy.
   (b) Number of placeable multidimentional requests on $h$ is given by $v_h = min(v_c, v_m.v_n)$

5. From $H$, find two hosts $h_{max}$ and $h_{s_max}$ which have maximum and second maximum number of placeable requests say $v_{max}$ and $v_{s_max}$ respectively.
6. If pair is found then
   (a) Increase number of placeable requests. $N_m \leftarrow N_m + v_{s_max}$.
   (b) Reduce the number of placeable requests of $h_{max}$ and $h_{s_max}$ by $v_{s_max}$ (i.e., $v_{max} \leftarrow v_{max} - v_{s_max}$ and $v_{s_max} \leftarrow 0$).
   (c) Remove $h_{s_max}$ from $H$.
7. If there are more than one host remaining in $H$ then goto step 5.
8. Possible number of placeable requests of last host, say $h_{last}$, denoted by $v_{last}$ is assigned as residual request capacity of the `reach` $r$. $r.res_req \leftarrow v_{last}$
9. $N_m$ gives the number of placeable requests.

It can be seen that procedure of finding the number of placeable requests $N_m$ inside reaches is similar to that of finding $T_R(nw)$. In Figure 6 if, instead of "available NIC bandwidth", "number of placeable requests" is used then it can be seen that the procedure is same. Similarly, the calculation of the placeable requests between `reaches` can be calculated in similar fashion as $T_{OR}(nw, b)$ with $r.res_req$ replacing $r.res_bw$.

### 3.4 Limitations of current network RRF Calculation

We discuss some of the limitations of currently proposed calculation of RRF and Fragmentation metric. The limitations are not in the idea or concept but in the ways they are calculated.

1. **Resource request representation for network RRF:** To calculate RRF, a representative request is required. For CPU, memory related RRF, the request which has been used in this paper is fine as request clearly represents a VM. However, to calculate network RRF, the request which has been used includes only two communicating endpoints. In reality the applications have more complex communication patterns as shown in [5]. Currently used request representation results in crude approximation of remaining capacity. The approximation can be improved by using more complex representation os requests like "Virtual Cluster" mentioned in [4] and "Tenant Application Graph TAG" mentioned in [13].

2. **Network RRF calculation:** Network RRF calculation procedure considers the request tuple to have same CPU and memory requirement for both the communicating end-points. Considering different requirements for end-points will involve more searching for suitable end hosts.

3. **Ignoring On-host network endpoints for network RRF calculation:** In the current RRF calculation, only the network capacity where the endpoints are at different hosts is considered even if the endpoints can be co-located on the same host. Current calculation in a sense gives the worst case remaining capacity for the requests.

## 4. FACTORS AFFECTING VM/APPLICATION PLACEMENT

This section begins by pointing out the important factors which affect VM placement. Then, the current related work in the field of VM placement is discussed in Section 4.1.

The two most important factors which dictate the choice of the VM placement schemes are

1. Application's resource requirements.

2. Data center network Architecture.

**Application's resource requirements:** VMs or applications need different resources like CPU, memory, NIC bandwidth, network communication bandwidth for execution. Different applications have different requirements. For example, applications like distributed DataBase and Kernel compile predominantly require CPU and memory whereas the applications like Key-Val store and Map-Reduce are more network intensive. This difference in the nature of resource requirements is crucial to decide the placement plan and hence which placement scheme to employ. The reason is that the resources such as CPU, memory and datacenter network bandwidth are inherently different from each other. CPU, memory and NIC bandwidth are kind of "Local" resources as they are shared by VMs on the same host. Core network bandwidth between communicating end points across the data center, on the other hand, is a "Global" resource as, theoretically, a link in data center can be shared by all the VMs hosted in the data center. Figure 7 shows the key differences between local and global resources. Resource conservation or VM placement strategies are different for local and global resources as mentioned in Table 1.

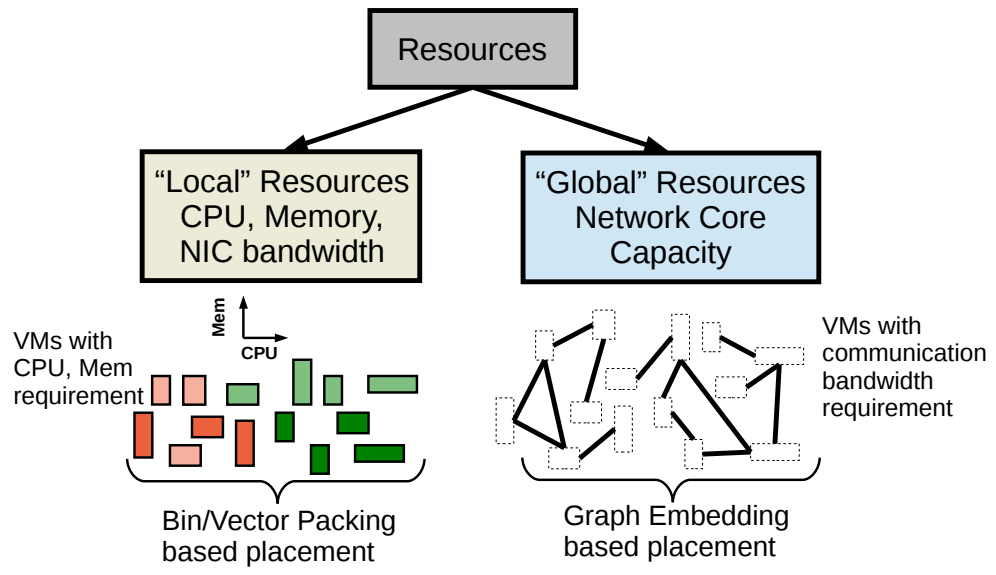

**Figure 7: Local vs Global Resources**

It can be seen that a unified scheme which optimizes resource utilizations across a set of inherently different resources can not be formed by simply optimizing one resource after the other in any sequential or parallel manner. These resources can not be optimized one after the other as assumed by [4, 13].

| | **Local Resources** | **Global Resources** |
|---|---|---|
| **Sharing** | **(a)** Shared / contended by VMs hosted on same PM, i.e., fewer contenders **(b)** identifying contenders is easy | **(a)** Potentially shared / contended by all the communicating VMs **(b)** identifying all the contenders is not easy |
| **Conserving / Optimizing heuristics** | Bin and vector packing heuristics are applied as resource capacities are properly defined [26] | Capacity of the network infrastructure depends on the way it is utilized. Graph Embedding heuristics are used [4, 13] |

**Table 1: Local Vs Global Resources**

**DC network Topology:** Other important factor to dictate the choice of VM placement scheme is the data center network architecture. The network infrastructure of data center can be "fat", i.e., providing high bisection bandwidth, or it can be "thin" like a tree which provides low bisection bandwidth. When the applications are network intensive, the choice of relative placement of communicating VMs directly affects the amount of bandwidth available between them. In low bisection bandwidth topology based data centers, the relative placement of communicating VMs becomes a critical issue. The same is not true when the data center topology has high bisection bandwidth. Infact, in case of full bisection bandwidth based topologies, like Fat Tree, the network bandwidth availability no longer remains a constraint. Paper [10] discusses and compares different data center topologies like CLOS, Fat-Tree and some non tree based topologies like Dcell, Full-Mesh. Due to the impact of the topology on the performance of Applications hosted on data center, different topologies, many of which are application specific, are proposed in [10, 23, 9, 8, 24].

### 4.1 Related Work

Most of the current VM/Application placement schemes are either local resource based placement scheme or network based placement schemes. Such schemes are not suited for applications which have mixed resource requirements. Table 2 shows the relative positioning of different schemes proposed in literature.

The performance of the mentioned schemes are good in their respective domains. The scheme proposed in this paper considers both local and network resources and hence can be used to place applications with mixed resource requirements.

[7] comes closer to our work in terms of number of resources considered for application placement. [7] considers a weighted function to combine CPU and network resources to utilize in placement decision making. They also consider a novel concept of "cold spots" which is a region in data center (potentially spanning across multiple racks) where CPU and network resources is relatively free. Unlike our work, this work doest not exploit topological property like `reach` and application requirements characteristics to placement decisions.

| App Res Req \ Reach Size | Large Reach | Small Reach |
|---|---|---|
| Predominantly local | Stochastic VM Multiplexing [19], Joint VM Provisioning [14], CBP, PCP based Placements [25], Sandpiper [26] | Same as Left |
| Mixed | Same as above | **VOID** |
| Predominantly network | Same as above | Traffic-aware VM Placement [15], Application-Driven BW Guarantees [13], Towards Predictable DC networks [4], Choreo: Network-Aware Task Placement [12] |

**Table 2: Relative positioning of Unified Resource Based VM placement Scheme**

Another body of work which is complementary to the placement scheme proposed in this Paper. That work tries to improve the performance of data center network after the placement has been done. For example Hedera [3], shows that TCP is agnostic about the multipath scenario of today's data center and proposes a dynamic flow scheduling system to efficiently utilize aggregate network resources. Other work deals with distribution of bandwidth among the VMs, like Seawall [21] which tries to provide a max-min fairness to the flows of different VMs. There are some Hypervisor based mechanisms like Gatekeeper [20] which provides hypervisor based rate limiting and feedback mechanism to avoid congestion on network links. A quality discussion about the desirable properties of schemes to improve performance of DC network is given in [18].

## 5. DESIGNING UNIFIED VM PLACEMENT SCHEME

In this section, the key insights regarding the datacenter topology gained in calculating the network RRF are used to design a unified application placement scheme. The goal is to have a placement scheme which, in terms of RRF of datacenter, performs atleast as good as already existing schemes (mentioned in Section 4.1) in their respective domains but significantly out-performs the current schemes when applications to be placed have mixed resource requirements. In other words, such a unified scheme should be capable of placing applications with any kind of resource requirements over any datacenter topology.

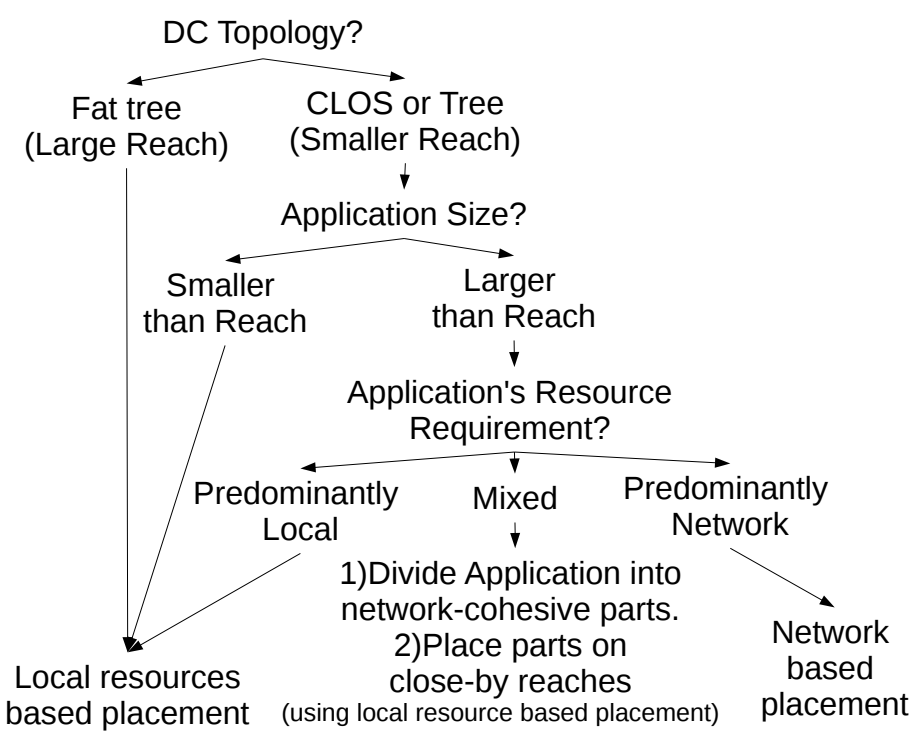

**Figure 8: VM placement decision tree**

The decision tree mentioned in Figure 8 is derived from the key insights gained in formulating the RRF metric calculation. Some important insights are:

1. Inside a `reach`, network bandwidth availability is not a constraint due to full bisection bandwidth, thus
   (a) Complex network based placement scheme can be avoided. Applying local resource based VM placement schemes are sufficient to gurantee resource availability.
   (b) Network can be treated as "local resource" by considering network requirements of communicating VMs only host's NIC rather than the whole path..
2. If possible whole application should be placed inside a single `reach`.
3. Between `reaches`, network bandwidth availability can be relatively constrained, thus
   (a) If application is hosted on multiple `reaches` then the division of application should be done to minimize inter-`reach` traffic.
   (b) Parts of application should be placed on close-by `reaches`.

The two points in the decision tree mentioned in Figure 8 which demand further explanation are 1)Application Size and 2)Application resource requirement characteristics. Size of an application can be represented in many ways. Some examples of application size are: a)number of VMs, b)aggregate resource requirement of applications VMs and c)number of communicating VM pairs. It can be seen that each of the size representation suits a different purpose. For network based VM placement schemes, "number of VMs" or "number of communicating VM pairs" may be more suitable representation of applications size. Similarly, for local resource based placement scheme, the "aggregate resource requirement" may be more appropriate representation of application size.

Defining application resource requirement characteristic poses a different kind of challenge. The question is how can an application be termed as more network intensive or more local resource intensive. The answers to these questions are most commonly given by looking at historical resource utilizations of the applications. Applications like Redis (key-val store) is considered to be more network intensive whereas distributed databases and kernel compilation are considered to be more local resource intensive. Resource utilization nature of Web applications and Map-Reduce kind of applications depends on the task they are executing.

## 5.1 VM Placement Algorithm

We now propose the unified VM placement algorithm. Following routines are used in the algorithm:

1. $Req(App\ a)$**:** Given an application $a$, $Req(a)$ denotes the 3-tuple $(c, m, n)$ which is representative size of VMs of $a$. A simple $Req(a)$ can be found out by taking the means of CPU, memory and network requirements of VMs.
2. $VMs(a)$**:** VM set of Application $a$.
3. $BW(X, Y)$**:** Bandwidth required between entities in set X and set Y.

**Algorithm 1** PlaceApplication($a$)

1: $V \leftarrow VMs(a)$
2: Let $r$ denotes the least loaded `reach`,i.e., the reach on which highest number of requests $Req(a)$ can be placed.
3: Let $V_r, V_{or}$ denote the subsets of VMs from $V$ currently placed inside and outside the reach $r$.
4: $V_r \leftarrow \varnothing$, $V_{or} \leftarrow V$
5: Choose VM $v$ from $V_{or}$ such that $BW(v, V_{or} - v)$ is highest
6: **while** $BAL_PACK(v, r) == SUCCESS$ **do**
7: $\quad V_r \leftarrow V_r \cup v, V_{or} \leftarrow V_{or} - v$
8: $\quad$ Find a VM $v_{next}$ from $V_{or}$ such that $BW(v_{next}, V_r) - BW(v_{next}, V_{or})$ is highest among all VMs in $V_{or}$.
9: $\quad v \leftarrow v_{next}$. Goto 6
10: **end while**
11: **if** $|V_{or}| > 0$ **then**
12: $\quad r \leftarrow BestSiblingReach(r)$
13: $\quad$ **if** $r! = NULL$ **then**
14: $\quad\quad$ GoTo 5
15: $\quad$ **end if**
16: $\quad$ Return FAIL
17: **end if**

We have used `BAL_PACK` [17] as the local resource based VM placement scheme. This scheme tries to maximize the number of VMs placed on given set of PMs such that the future resource shortfalls experienced on the PMs are below a certain pre-specified threshold. The idea is to avoid VM migrations happening because of resource shortfalls. The scheme utilizes the historical dynamic resource requirement data of VMs.

## 6. EVALUATION OF UNIFIED SCHEME

We evaluate the performance of the proposed VM placement scheme and compare it with existing local and network based placement schemes in terms of the resulted RRF using a datacenter simulator. Figure 9 shows the overall evaluation space. There are different DC topologies with different sized `reaches` possible. Resource requirements of Applications also vary from being predominantly local to predominantly network. Though evaluation space is large there are certain reductions possible. They are listed below:

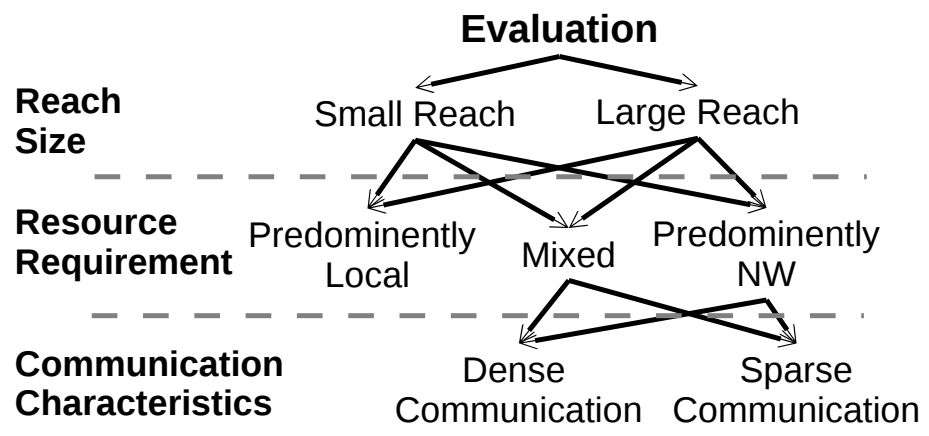

**Figure 9: Evaluation Space**

1. If a network based placement scheme performs well for small `reach` sizes, i.e, oversubscribed (low bisection bandwidth) DC topologies then it will perform well for non-oversubscribed DC topologies too. Thus, we use only simple Tree and CLOS based topologies for evaluation.

2. If a placement scheme performs well for applications with mixed resource requirement and dense communication pattern then it will perform well for applications which have predominantly network requirement. This is because the added constraints of fulfilling local resource requirements are not there. Thus for mixed resource requirement, evaluating the dense communicating scenario is sufficient to show effectiveness of the proposed scheme.

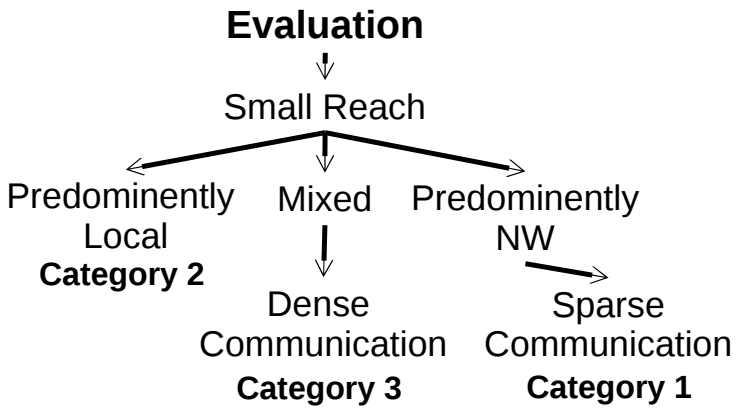

**Figure 10: Reduced Evaluation Space**

After reductions there are effectively only 3 cases to be considered as shown in Figure 10.

### 6.1 Base Case Schemes

We compare the proposed VM placement scheme named `UNIFIED` with the following local resource based and network based VM placement schemes.

`LOCAL`**: Stochastic VM Multiplexing [19]:** In this local resource based VM placement scheme first, the size (resource requirements) of the VMs are approximated stochastically from their historical resource utilization data. The most dominant local resource is used for the size approximation. Then VMs are placed on the PMs using this stochastic representative size using FFD based bin packing heuristic.

`NETW` **:Towards Predictable DC [4]:** The network based VM placement scheme mentioned in this paper uses a "hose" based application communication model called Virtual Cluster. This scheme assumes the structure of datacenter topology to be of a simple tree. Each VM is assumed to occupy identical slots in PM. Thus, scheme does not consider the local resource requirement of the VMs. To place an application, the scheme does a node by node scanning beginning with the lowest level, i.e. hosts. If a host has sufficient number of slots to place the application, then, VMs are placed on it. If no host has required number of slots then the scheme goes a level up in topology hierarchy. Scanning goes on till a suitable node which has enough number of available slots is found. The scheme is a First Fit scheme. The VC representation of the application considers all VMs to have same bandwidth requirement till the Virtual Switch.

### 6.2 Datasets Used

Table 3 mentions the datasets and the datacenter configuration used for experiments. Dataset [2] is available in form of TCP dump of LAN traffic consisting of mail servers, web servers which can be used to find communicating entities and their bandwidth requirements. Dataset in [5] is in form of a communication matrix between different services in Bing data center. There is no time varying information. However, mean communication bandwidth between entities can be approximated. Both of these datasets do not provide CPU and memory utilization information and we have approximated them for experiments. The approximation consisted of weighted sum of both incoming and out going network traffic and a uniformly distributed random number. The third dataset is synthetic.

### 6.3 Evaluation Steps

We compare the performance of `UNIFIED`, `LOCAL` and `NETW` schemes in terms of the network RRF they result in the data centers. Here are the steps used:

1. Randomly shuffle the set of applications to mimic an *on-line* placement scenario. The shuffled list of applications is same across compared placement schemes.

| Category | Topology | Host Cfg | Dataset | VMs/App | VM Cfg (mean req) | RRF calculating req |
|---|---|---|---|---|---|---|
| 1- Netw | Tree,64Hosts,10Gbps | 4 Ghz, 8 GB | Univ.[2] | 2 to 14 | 400Mhz,200MB,193Mbps | 400Mhz, 200MB, 200Mbps |
| 2- Local | CLOS,64Hosts,5Gbps | 8 Ghz, 16 GB | Bing[5] | 2 to 18 | 620Mhz,438MB,225Mbps | 600Mhz, 400MB, 200Mbps |
| 3- Mixed | CLOS,64Hosts,10Gbps | 8 Ghz, 16 GB | Synthetic | 10 to 15 | 500Mhz,700MB,100Mbps | 500Mhz, 700MB, 200Mbps |

**Table 3: Datasets used for experiments**

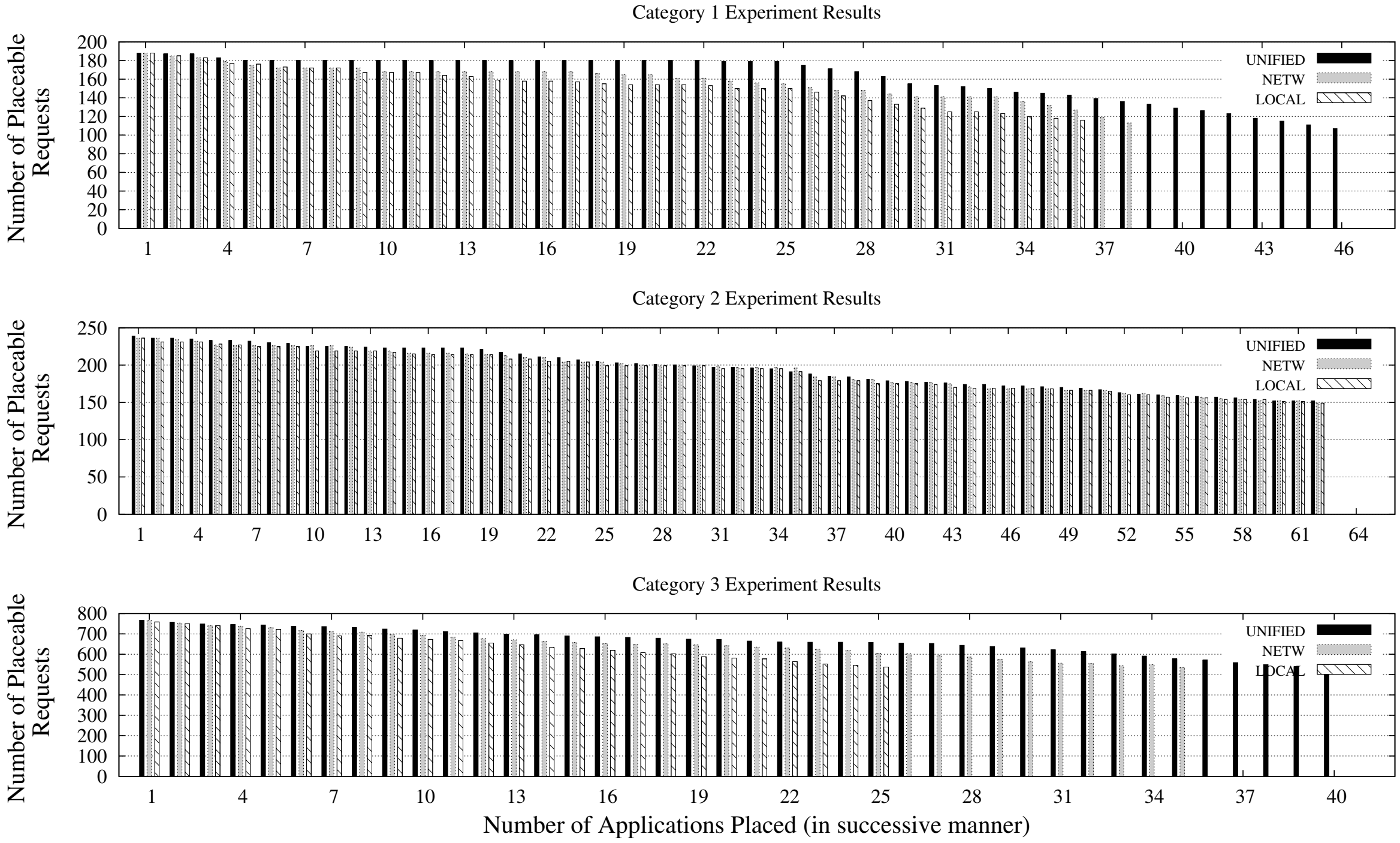

**Figure 11: Number of Applications placed vs datacenter RRF**

2. Place applications one by one till the time no more applications can be placed and after every successful placement calculate the RRF of the Datacenter. The RRF is represented in terms of number of placeable requests.

## 6.4 Results and Observations

Figure 11 shows the RRF levels experienced when different application placement schemes are used under different categories of experiments as mentioned in Table 3. The X-axis denotes the number of applications placed sequentially (one after the other) and the Y-axis shows the resultant RRF after every successful application placement. For example, the value 10 on the X-axis denotes that 10 applications have been placed and the corresponding Y-axis value shows the resultant RRF values.

**Observation 1:** For categories 1 and 3 where applications need predominantly network resource and mixed resources respectively, `UNIFIED` scheme is able to place 15 more applications than `LOCAL` scheme for experiment category 3 and 8 more applications than `NETW` for experiment category 1. Also, the number of placeable requests resulted by `UNIFIED` scheme is 20% higher than `LOCAL` and 12% higher than `NETW` after placement of 23rd application in category 1. The main reason for better performance of `UNIFIED` scheme is intelligent distribution of application's VMs across `reaches`. `UNIFIED` scheme utilizes local `reach` Capacity first and then only the higher links.

**Observation 2:** Surprisingly, for experiment category 2, where application requirements are predominantly local resources, there is hardly any difference in the performance of different placement schemes. The main reason is that for this dataset dynamic VM resource requirements are not available. Because of such static dataset, the placement process of applications got reduced to avoiding capacity violation on hosts. One more reason is that the inter VM network requirement is also highly skewed as shown in Figure 12 thus, the network links remain unutilized and network based schemes did not have any optimizations to perform.

# 7. CONCLUSION AND FUTURE WORK

In this paper we proposed a comprehensive metric, called RRF, for capturing the resource fragmentation

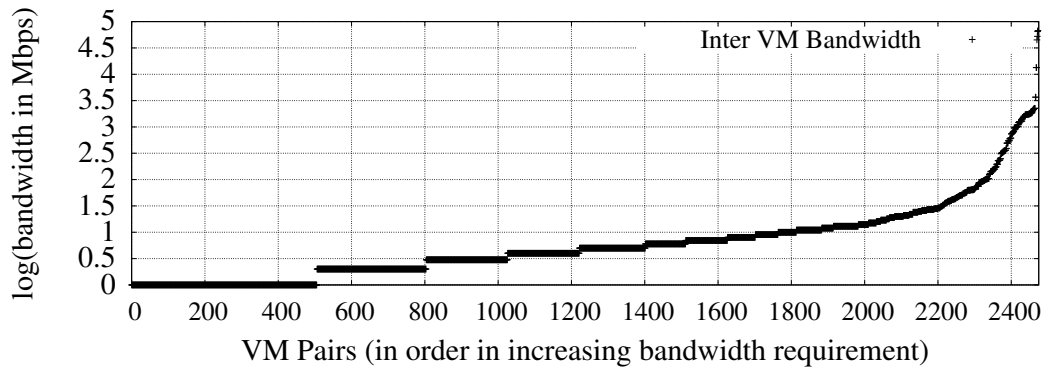

**Figure 12: Bandwidth requirement between VMs for category 2**

in a datacenter. We also proposed the novel concept of `reach` which can be used to find the achievable resource capacity of the datacenter. We used the concept of `reach` to design a unified application placement scheme which considers all the resources for planning the placement and thus reduces the RRF in datacenter. We evaluated the proposed application placement scheme and showed that it significantly improves the application hosting capacity of datacenter by reducing the resource fragmentation. The proposed work has other possible uses which have not been discussed in this paper. For example, the concept of `reach` can be used by a cloud provider to find out the most suitable VM instance size to offer so that the fragmentation in datacenter is minimized. Similarly, by using the concept of RRF, a cloud provider can identify the bottleneck resource in datacenter. In future we want to extend this work to include

1. Finding `reaches` in non-tree based topologies.
2. Evaluation of placement scheme on non tree based topologies.
3. Representing network RRF request in a more complex form like a Virtual Cluster [4] or TAG [13].
4. RRF calculation procedure considering different resource requirements for request end-points.